\documentclass[11pt,twoside]{article}
\usepackage{newpasp}

\def\edcomment#1{\iffalse\marginpar{\raggedright\sl#1\/}\else\relax\fi}
\reversemarginpar

\begin{document}
\title{Theory of Pulsar Winds}
 \author{Jonathan Arons}
\affil{University of California, Berkeley, Department of Astronomy,
601 Campbell Hall, Berkeley, CA, 94720-3411, USA}

\begin{abstract}
I discuss recent work on the nature of relativistic winds from
Rotation Powered Pulsars, on the physics of how they transport
energy from the central compact object to the surrrounding world, and 
how that energy gets converted into observable synchrotron emission.
\end{abstract}

\section{Introduction}

Rotation Powered Pulsars (RPPs) are the prime examples where the electromagnetic 
extraction of energy from a rotating compact object clearly provides
the power for the nonthermal photon emission that distinctively 
characterizes the world of High Energy Astrophysics. While the fact that
the energy extraction reflects electromagnetic torques exerted on a
neutron star by the macroscopic electromagnetic fields with which each
star is endowed - indeed, the magnitude of that torque allows one to infer
the strength of each star's magnetic moment - has been known since the 
earliest days of pulsar research (Gold 1968), and indeed was predicted
(Pacini 1967) even before pulsars were discovered, 1) the physics of the
processes through which the extraction works, 2) the physics of how the 
rotational energy is transmitted to the surrounding world, and 3) the
physics of how that energy transforms into the observed synchrotron 
radiation from the nebulae around pulsars (when these are observed) have all
remained open questions through more than 30 years of pulsar research.
Answers to all three questions are of significance not only
to the understanding of RPPs themselves, but also to the physics of
Active Galactic Nuclei, especially issues of jet formation and blazar
emission, and to the workings of Gamma Ray Burst sources, especially
to those issues involving the physics of relativistic shock waves and
possibly to the nature of the basic engine underlying the GRB phenomenon. 

The earliest theoretical answer to 1) and 2) was the vacuum wave theory of
electromagnetic spindown, a theory which has never satisfactorily led to
useful answers to 3).  Modern pulsar theory suggests the answer to 2) is 
that a RPP throws off its rotational energy in the form of a relatively
dense magnetized, relativistic wind of plasma (Michel 1969), largely composed of
free electron-positron pairs with an embedded, wound up magnetic field. 
Support for this idea has always come from X-ray astronomy.
Observations of course provide the basic input to all attempts to model
these phenomena. Recent 
advances  have been especially powerful drivers of progress on question 3) 
and to some aspects of question 2). 

Study of the Crab Nebula, whose Chandra X-ray picture (Weisskopf {\it et al.}
2000) appears in Figure 1, and more recently of other young 
pulsar wind nebulae (PWN), has been the source of most of the 
conceptual machinery in this field.

The most widely accepted hypothesis for understanding what we see is that 
the observed nebular emission is synchrotron radiation from particles 
heated into a nonthermal distribution of energies by the termination
shock waves in the magnetized relativistic $e^\pm $ pair outflow(s) from
the underlying pulsar. My focus (obsession, if you like) is on the physics of 
these shock(s) and what that has to to tell us about the wind powering the 
shock. The observations (including optical observations first done by 
Lampland 1921) demonstrate the intrinsic time variability of this shock 
transition. Therefore, time dependent modeling of the dynamics of these
relativistic flows is needed, with results which yield quite interesting
constraints on the central engine, as well as telling us a lot of 
interesting things about the 
shocks themselves, conclusions of relevance to AGN and GRB as well as to 
pulsars and the PWN.

One important fact must be kept in mind in interpreting these  
rather different looking systems. The Crab is a relatively
compact, high magnetic field nebula.  At X-ray emitting energies, the
radiating particles lose their energy in approximately a flow time from the
termination shock in the pairs (perhaps at or just inside the inner X-ray
ring) to the outer edge of the visible X-ray torus, if the flow
velocity is that of the Rees and Gunn (1974) model (recovered in the
MHD model of Kennel and Coroniti 1984a), $ v = (c/3) (R_{shock}/r)^2 $. 
This strong cooling is behind the well know fact that the nebular image
contracts with increasing photon energy from infrared through 10 keV
X-rays - interestingly, what little is known of the nebular size at higher 
photon energies (Pelling {\it et al.} 1987) suggests that nebular 
image contraction may come to an end 
in the 100 keV - 1 MeV range, which if true would suggest that
the shock heating = ``acceleration'' is distributed over an observationally 
resolvable region.

Other nebulae are not always so radiatively efficient. G320.4 around 
PSR B1509-59 clearly has a much weaker magnetic field, doubtless reflecting
different confinement properties of its local interstellar medium, with
synchrotron losses affecting the spectra of outflowing particles only at
much larger radii, if at all (Gaensler {\it et al.} 2002), even at X-ray 
emitting energies. 3C 58 also seems to have rather low radiative efficiency.

The nebulae's appearance shows energy injected in equatorial outflow and 
polar jets. The ratio of polar to equatorial energy flux is not known, nor
do we know whether the observed brightness distributions, including the
apparent angular gaps between equatorial torii and polar jet, reflect
gaps in the energy injection or changes in the energy conversion processes,
such that mid latitude energy injection is darker than polar and equatorial
phenomena.

\section{Energy Flow}

However, especially in the case of the Crab, the detailed observations 
accumulated over many years tell us a lot. The optical polarization in the
nebula's inner minute of arc tells us the magnetic field is toroidally
wrapped around the Nubula's long axis, which is pretty well co-aligned 
with the pulsar's rotation axis (Rees 1972, using Woltjer's 1958 data.)
The equatorial wind's global energy conservation allows us to define many
of the concepts essential to current discussion. According to the 
gospel of Kennel and Coroniti (1984a), the pulsar's rotational energy 
gets carried away in a relativistic magnetohydrodynamic outflow composed of 
toroidally wound up magnetic field frozen to an electron-positron-heavy 
ion plasma - curiously, the heavy ions, a feature not in the original 
Kennel and Coroniti model, remain controversial, although in fact the 
justification for the pairs is only that theorists cannot think of any 
other way a pulsar might supply the particle injection rate needed to
continuously feed the nebular X-ray source - 
$\dot{N}_\pm \sim 10^{38.5-39}$ particles per second for the Crab, much 
in excess of the fiducial Goldreich-Julian elementary charge loss rate
$\sim 10^{34.5}$ s$^{-1}$ for this pulsar.  As has been argued some time
ago, a detailed kinetic (non-MHD) look at the equatorial wind's 
termination region suggests the presence of a Goldreich-Julian heavy ion 
current in the equatorial wind (Hoshino {\it et al.} 1992, Gallant and 
Arons 1994), a topic to which I return below.  

For a steady wind outflowing radially in a spherical sector opening
into a solid angle $\Delta \Omega$, energy conservation reads
\begin{eqnarray}
     \dot{E}_R  & = & r^2 \Delta \Omega \left\{ c\beta 
      \frac{B^2}{4\pi} + c\beta \gamma \left[(n_+ + n_- ) m_\pm c^2
      + n_i m_i c^2 \right] \right\}  \nonumber \\ 
   &  = & 4\pi r^2  \frac{\Delta \Omega}{4\pi} c\beta \frac{B^2}{4\pi} 
       \left(1 + \frac{1}{\sigma}\right),\label{eq:energy-cons}  \\
   \sigma & \equiv & \frac{B^2}{4\pi \rho \gamma c^2}, \, 
   \rho \equiv (n_+ + n_- ) m_\pm +  n_i m_i, \, \gamma \gg 1. \nonumber
\end{eqnarray}
The ideal MHD theory of Kennel and Coroniti tells us to expect
$rB = $ constant, $r^2\rho =$ constant, $v = {\rm constant} \approx c$, 
$\gamma = $ constant and therefore $\sigma =$ constant in the freely 
expanding wind. Everything known or thought about the pulsar says that where
the wind is ``launched'' ($r \approx R_L \equiv cP/2\pi $, the light cylinder), 
$\sigma \gg 1$. In the outer magnetosphere and inner wind,
the energy is tied up in nonradiating (in the sense of 
directly observable electromagnetic radiation) EM fields and plasma flow -
``the dog doesn't bark in the night'' (Arons 1992). Recent HST (Hester
{\it et al.} 1995) and Chandra (Mori, these proceedings) observations
show various small ``knot'' and ``sprite'' features that may be radiative 
signatures of the expanding wind, but little is clear in this subject.

If $\sigma \gg 1$ everywhere in the wind, as it would be if ideal MHD were
everywhere the rule, expression (\ref{eq:energy-cons}) yields the traditional 
estimate, going all the way back to vacuum wave theory, for the wind's  
magnetic field at the radius where the dynamic pressure of the wind is 
comparable to the total pressure of the nebular bubble,
$B_1 (\sigma \gg 1) = (4\pi p_{nebula})^{1/2} = 350 \;\mu$G; the numerical
value applies to the Crab Nebula. A comparable estimate for G320.4 yields 
a little less than 11 $\mu$G for $B_1$. These values can be fudged to be comparable 
to the numbers found by applying equipartition estimates to the nebulae.
Therefore, in the earliest days of wind theory, everyone felt comfortable
with this admittedly crude modeling. Better understanding of the nebular 
dynamics, in the context of one dimensional flow models, soon disrupted
any sense of comfort.

\section{Pulsar Wind Nebulae as $\sigma_1$ Probes}

Pulsar Wind Nebulae = PWN, formerly known as Plerions 
(Weiler and Panagia 1978 - the
name referred to the closed nebulae that form when the pulsar's space
velocity is negligible), are, 
in the context of the young and strongly driven systems of interest here,
thought to be light weight bubbles confined by inertia or 
ISM/CSM pressure (ram and static) (Rees \& Gunn 1974, Kennel \& Coroniti 
1984a), fed by the particles and fields from the pulsar. Schematically, 
their structure is as in Figure 2.
They have turned out to yield surprising information on the wind, by 
providing constraints on $\sigma_1$, the ratio of electromagnetic energy 
flux to the plasma kinetic energy flux in the wind just interior to the
shock wave thought to terminate the wind's free expansion. In the case of 
the Crab Nebula, this shock must occur at roughly $10^9 R_L$; the shock
radius is also very large, in terms of light cylinder radii, for all the
other PWN studied to date. The surprising 
result, obtained from several points of view, is that $\sigma_1 \ll 1$;
typical values obtained have been less than $10^{-2}$.
Somehow, between the light cylinder and the wind termination shock, 
magnetic energy has disappeared without reappearing to us as observable 
photons, even though back at the light cylinder the outflowing magnetic 
(and electric) field carried most of the pulsar's rotational energy. This
apparent disappearance of large scale electromagnetic energy in favor
of bulk flow kinetic energy, with little visible emission, is what is 
now called ``the $\sigma$ problem.''

The evidence for low $\sigma$ comes from three model dependent sources -
modeling the bulk dynamics of the synchrotron emitting bubble (the visible
PWN); modeling the radiative emission of the PWN; and modeling the 
collisionless shock wave that mediates the transition from invisible
wind to visible bubble.

\subsection{Estimating $\sigma_1$ from PWN expansion velocities}

Empirically, observed PWN expand non-relativistically.  The MHD jump
conditions for the relativistic shock that terminates the wind show that
nonrelativistic flow of the lightweight fluid (magnetic field and 
relativistic plasma emerging from the shock) only if $\sigma_1 \ll 1$ 
(Kennel and 
Coroniti 1984a, Gallant {\it et al.} 1992), at least so long as the flow is 
laminar. If the 
magnetosonic speed is that of a relativistic plasma throughout the bubble,
the implications of the jump conditions are preserved (Rees and Gunn 1974,
Kennel and Coroniti 1984a), even when the time dependence of the expansion
is included (Emmering and Chevalier 1987) and when the one dimensional flow
assumption is dropped (Begelman and Li 1992). All these models lead to
$\sigma_1 \approx 0.003-0.005$ for the equatorial wind in the Crab. 
Therefore, the magnetic field just inside the wind's termination shock 
(treating that shock as having infinitesimal thickness) yields
\begin{equation}
B_1 = \left(\frac{\sigma_1}{1 +\sigma_1} \right)^{1/2} B_1(\sigma_1 \gg 1)
    = \left(\frac{\sigma_1}{1 +\sigma_1} \right)^{1/2} (4\pi 
    p_{nebula})^{1/2} =  19 \; \mu {\rm G}.
\end{equation}
The numerical value applies to the Crab. The analogous value for G320 is
0.8 $\mu$G (Gaensler {\it et al.} 2002).

Such models have the nice feature of having a decelerating radial flow of
the plasma in the bubble, $v = (c/3)(R_{shock}/r)^2$, which in turn leads to
the magnetic field in the bubble increasing linearly with radius,
$B = 3 B_1 (r/R_{shock})$ until it
reaches approximate equipartition at $R_{eq} = (2/9\sigma_1)^{1/2} R_{shock}$. 
For $r > R_{eq} $, deceleration stops. Matching the asymptotic bubble flow 
velocity to the observed bubble expansion speed (when that can be measured 
or inferred) is what leads to the inferred low $\sigma_1$.

The quantitative model which allows one to draw this inference assumes 
toroidally wound $B$ field (as suggested by the theory and by nebular
polarization observations), a lightweight fluid composed of pairs and $B$ 
field. Like all starting models, this one leaves lots unanswered. Where 
has all the upstream $B$ field gone? How come the inferred upstream flow
4-speed $c\beta \gamma, \; \sim 10^6c$ in the Crab, is so small compared to 
the natural acceleration scale 
$\sim e\Phi_{open} /m_\pm c^2 \approx 10^{11}$; the numerical value is for 
the Crab? The choice of $R_{shock}$ is unclear - locating the shock in the 
pairs at the
inner X-Ray ring in the Chandra imaging puts the equipartition radius in
the midst of the X-ray torus, which is surprisingly small. 

\subsection{Estimating $\sigma_1$ from radiation modeling of the Bubble}

One can also extract $\sigma_1$ by adding a couple of extra bells to the 
model's whistle. If one assumes the flow energy goes promptly into power law
distributions of pairs right at the shock, via ``shock acceleration''
(a.k.a. magic), the MHD jump conditions give a lower cutoff to the power
law of $\gamma_{min} \approx 0.6 \gamma_{wind}$, assuming a 3D plasma with
ratio of specific heats equal to 4/3. With the spectral slope of the 
particle distribution as an adjustable parameter, adequate modeling of the 
high energy photon (optical and harder) spectrum of the Crab can be achieved 
(Kennel and Coroiniti 1984b), with a result for $\sigma$ similar to that 
obtained from dynamical arguments. One finds $\sigma_1 \approx 0.005$, 
$\dot{N}_\pm \approx 10^{38.5} \; {\rm s}^{-1} = 10^4 \dot{N}_{GJ}$, where
$\dot{N}_{GJ} = 2c\Phi_{open}/e $ is the Goldreich-Julian particles/charge
outflow rate and $\Phi_{open} = (\dot{E}_R /c)^{1/2} $. Interestingly,
the inferred pair injection rate is similar to that found in ``polar
cap'' models of pair creation, which yield 
$\dot{N}_\pm /\dot{N}_{GJ} \equiv \kappa_\pm \approx 4 \times 10^4 $ 
({\it e.g.}, Hibschman
and Arons 2001) for this pulsar. Standard theoretical scenarios suggest
these pairs expand to fill all $4\pi$ sterradians around the pulsar, so
the factor of 4 discrepancy may reflect no more than the special conditions 
needed to make the equatorial part of the wind shine brightly, to which I
return below. 

In the Crab, the fact that the synchrotron cooling time of the
accelerated particles that emit the X-rays is comparable to the flow
time makes extracting these parameters from radiation models a relatively
straightforward issue. Extracting similar parameters for Vela and G320.4, 
for example, is harder and dependent on more sophisticated modeling, since
$t_{flow} \gg t_{synch}$.  Nevertheless, progress is possible, based on
a more sophisticated strategy, with conclusions similar to those from the
Crab: $\sigma_1 \ll 1, \; \dot{N}_\pm \approx 
\kappa_\pm ({\rm polar \; cap \; theory})  \dot{N}_{GJ}$. I don't have space 
to repeat those arguments here; suffice to say, that properly applying the
Kennel and Coroniti model to the recent Chandra observations of the Vela 
nebula (Helfand {\it et al.} 2001) leads to the conclusion that 
$\sigma_1 < 0.05$, rather than $\sim 1$ - Helfand {\it et al.} assumed a
constant velocity rather than a decelerating flow downstream of the wind 
termination shock.

\subsection{Estimating $\sigma_1$ from the Kinetic Theory of the
Termination Shock: Wisps as Ion Driven Compressions}

Driven by curiosity as to just how a highly relativistic, collisionless 
shock wave in a magnetized medium, with the $B$ field almost transverse
to the flow, goes about transforming upstream flow energy into the 
observed distributions of synchrotron emitting particles, Hoshino {it et
al.} (1992) and Gallant and Arons (1994) introduced a theory of the
termination shock structure which appears to be of some use in the 
interpretation of the features seen in the spectacular movies and images of
the Crab and other young PWN coming from the HST and Chandra. Dissatisfied
with the oft repeated assertion that ``shock acceleration'' (meaning 
diffusive Fermi acceleration) in a 
highly relativistic flow actually works
as advertised when the magnetic field is not almost strictly parallel
to the flow velocity, these authors found instead that non-thermal
acceleration of $e^\pm$ pairs actually might work as a
consequence of the flow having much of its energy carried in heavy ions
(``protons'').  Then the shock structure starts with a thin
(shock thickness $\delta \sim$ pair Larmor radius) transition
in which the pairs are heated to a downstream {\it thermal} distribution
with temperature comparable to $\gamma_1 m_\pm c^2 $. The heavy ions 
penetrate this hot magnetized $e^\pm $ plasma, and begin gyrating in the
pair shock compressed magnetic field. Such gyration is electromagnetically
unstable, with the coherent gyration of the ions degenerating as they
collectively emit low frequency electromagnetic waves (technically,
magnetosonic waves) in the pair plasma. Resonant cyclotron absorption of
these waves by the pairs leads to pregressive nonthermal acceleration
of the pairs as they flow away from the shock front.

The details of this acceleration scheme still need further development. 
What has proven to be useful is the observation that the ions' gyration
causes the deposition of the ions' outflow momentum into a series of 
compressions of the magnetized pair plasma, in the form of a more or less
standing wave whose compressional oscillations emit traveling waves that 
move into the plasma at larger radii. These features have more than a 
little resemblance to the X-ray ring and arcs seen in the Crab and G320.4,
and to the wave like features seen in the HST movies of the Crab Nebula, as
illustrated in Figure 3, which comes from hybrid (MHD flow of
pair plasma and Particle-in-Cell treatment of the ions) simulation of
the termination shock region (Spitkovsky and Arons 2000, and submitted
to ApJ), a region thought to include the moving ``wisps'' in 
the Crab.

Rough comparison of the synthesized moving features in the surface brightness 
to the observations yields $\sigma_1 \approx 0.005, \; \gamma_1 \approx
3 \times 10^6, \; \dot{N}_\pm \approx 3 \times 10^{38} \; {\rm s}^{-1}, \;
\dot{N}_i \approx 2 \times 10^{34} \; {\rm s}^{-1} \approx \dot{N}_{GJ}.$
However, serious comparison to the temporal structure observations, now 
much refined by recent HST and Chandra campaigns (Hester, these proceedings;
Mori, these proceedings) will require lifting the spatially 1D assumptions 
underlying this model. However, it is interesting that the same ideas can
be usefully used to describe the inner arc features seen in the single 
Chandra snapshot of G320.4 (Gaensler {\it et al.} 2002; Gaensler {\it et
al.}, these proceedings.)

Most other ideas about the wisp region in the Crab say nothing as definite 
as is in the model just outlined.  
Begelman (1999) suggested the wave like features might reflect a 
Kelvin-Helmholtz instability between equatorial and higher latitude flows 
outflowing with different velocities, an idea which may well have merit
in the face of the recent observations - it certainly deserves more 
quantitative development. Hester (1998) has been more specific, in
suggesting the wisp features to be compressions comoving with the flow, 
created by thermal instability driven by synchrotron cooling. This idea
might work in principle, if the X-ray emission in the wisp region has a
sufficiently flat spectrum, for then the particles carrying the pressure
are subject to rapid synchrotron cooling - a particle spectrum flatter
than $E^{-2}$ is required.  However, this explanation does not apply to 
the ring and arc features in the Vela and G320.4 nebulae, where the 
synchrotron time clearly is much larger than the flow time, a difficulty
not a problem for the ion doped wind model's relevance to these systems.

Lest one think theories of pulsar winds and their interaction with the
surrounding nebulae are anything like complete, recent radio observations
by Bietenholz {\it et al.} (2001) strike a cautionary note. All the physical
models described to date yield lower cutoffs to the power law distributions
of radiating particles at energy $\sim 0.5 \gamma_1 m_\pm c^2 $.  The
inferred value of $\gamma_1$ in the Crab says there should be no observable
radio wisps. The observations shown in Figure 4 say 
otherwise. There is no ready explanation at hand for this fact - some deeper
thinking about the shock acceleration problem (if indeed any kind of shock
actually does terminate the wind) is needed.

There have been various attempts to wiggle out of the ``sigma problem'' - 
e.g., Begelman's (1998) invocation of MHD kink instabilities and 
reconnection as a means of turning the stiff ordered toroidal $B$ field
into a softer gas of magnetic cells 
($B^2 /8\pi \propto ({\rm density})^{4/3}$) - such a soft gas model 
leads to inferring $\sigma_1 \sim 1$ (not $ \gg 1$, however.) Whatever one
may think of the physics of such speculations, the high optical polarization
in the Crab's inner regions forbids such cellularization
of the $B$ field. Squirting the gas of an equatorial MHD nozzle at the 
light cylinder (Chiueh {\it et al} 1998) seems to have little to do with
what little we understand of the dynamics of the flow at the light 
cylinder, and putting all the flow in polar jets (Sulkanen and Lovelace
1990) violates the obvious fact that most of the energy lost seems to
be in the equatorial wind.  Dissipation of the wind, considered as wound up
magnetic sectors (Coroniti 1990) seems to be too weak (Lyubarskii and Kirk
2001). 

Most of the difficulties come from theorists' consideration only
of an unstructured
wind in ideal MHD. Coroniti's (1990) model (and Michel's original 1971
suggestion) points the way to consideration of winds with wave like
structures, which must be present in outflows from oblique rotators, as has
been emphasized most recently by Melatos (1998). Perhaps a theory of
winds plus strong electromagnetic waves in the outflowing plasma will do
better at resolving the origin of small $\sigma_1$, as well as giving us an
idea of what leads to polar ``jet'' formation. In this regard, the very
preliminary results shown at this meeting by Spitkovsky may be a harbinger 
of things to come.

\acknowledgments

What I have learned about pulsars and pulsar wind nebulae over the years 
has benefitted from more than a little help from my friends: David Alsop, 
Elena Amato, John Barnard, Bill Fawley, Bryan Gaensler, Yves Gallant, Johann 
Hibschman, Masahiro Hoshino, Vicky Kaspi, Bruce Langdon, Claire Max, Ted 
Scharlemann, Anatoly Spitkovsky and Marco Tavani.  I also acknowledge the
(currently interrupted) support of the US National Aeronautics and Space
Administration and of the National Science Foundation, and the never failing
support of California's taxpayers.

\newpage

\noindent Figure Captions
\newline

\noindent Figure 1: a) The Crab Nebula (age = 947 years) in X-rays, 
with equatorial 
X-ray torus and inner  X-ray ring, plus polar ``jets''; b) G320.4 around 
PSR B1509-58 (spindown age = 1250 years), with a 
partial X-ray ring around the pulsar and a (visibly) one sided X-ray jet
(Gaensler {\it et al.} 2002); c) 3C 58 in X-rays, with age 815 years if 
this is the remnant of SNe 1181, with $T_{spin}$ of the X-ray pulsar much 
larger, 3335 years; d) the much older ($T \sim 10^4$ years) Vela PWN  years in 
X-rays. 
\newline

\noindent Figure 2: Schematic of a confined PWN, for which the pulsar
space velocity is small compared to the bubble's expansion speed. 
\newline

\noindent Figure 3: Top Row: Observed time series of the wisp region in the Crab 
Nebula (Hester {\it et al.} 1995). The physical scale of the separation 
between the pulsar and the main bright wisp to the pulsar's northwest
(the first feature concave to the pulsar) is $7^{\prime \prime}$, or
0.15 pc. Middle row: Pair fluid is ``stirred'' by the instability of the ion
stream entering the shock front and executing Larmor gyration in the compressed
magnetic field.  Each panel shows a snapshot, taken at successive times,
of the magnetic field
(top curve), ion momentum (middle curve) and ion density (bottom curve), each as 
functions of radius, with the radii measured in units of the ions' Larmor radius
based on parameters of the upstream flow ($\sim 0.15$ pc in the specific case
shown.) New waves are emitted with every turn of the bunched ``knot''
on the first loop in the ion orbit. Bottom row: Synthetic surface brightness maps,
corresponding to each snapshot.
\newline

\noindent Figure 4: VLA image of the radio wisps around the Crab pulsar (Bietenholz
{\it et al.} 2001), which suggests
the radio emitting electrons in the Crab Nebula have the same source as the
particles emitting the harder photons.

\end{document}